\documentclass[aps,twocolumn,superscriptaddress,footinbib]{revtex4-1}

\usepackage{longtable}
\usepackage{morefloats}
\usepackage[dvips]{graphicx}
\usepackage{color}
\usepackage{epsfig,graphicx,amsfonts,amsbsy}
\usepackage{amsmath,amsfonts,amsthm,amssymb}
\usepackage{appendix}
\usepackage{makeidx}
\usepackage{url}
\usepackage{verbatim}
\usepackage[bookmarksnumbered,pdfpagelabels=true,plainpages=false,colorlinks=true,linkcolor=blue,citecolor=red,urlcolor=blue]{hyperref}
\usepackage[rightcaption]{sidecap}
\usepackage{array}
\usepackage{multirow}
\usepackage{tabularx}
\usepackage{braket} 
\usepackage{dsfont}
\usepackage{bbold}
\usepackage{etoolbox}

\usepackage{dcolumn}
\usepackage{bm}
\usepackage{blindtext}
\usepackage{comment}

\usepackage{newtxtext}
\usepackage{newtxmath}

\def\CalE{{\mathcal{E}}}

\usepackage{xcolor}
\usepackage[normalem]{ulem}

\begin{document}

\title{Data-driven reconstruction of spectral conductivity and chemical potential using thermoelectric transport properties}

\author{Tomoki Hirosawa}
\affiliation{Department of Physics, University of Tokyo, Bunkyo, Tokyo 113-0033, Japan}
\affiliation{Department of Physics, University of Basel, Klingelbergstrasse 82, CH-4056 Basel, Switzerland}
\affiliation{Department of Physical Sciences, Aoyama Gakuin University, Sagamihara, Kanagawa 252-5258, Japan}

\author{Frank Sch\"{a}fer}
\affiliation{Department of Physics, University of Basel, Klingelbergstrasse 82, CH-4056 Basel, Switzerland}
\affiliation{CSAIL, Massachusetts Institute of Technology, Cambridge, MA
02139, USA}

\author{Hideaki Maebashi}
\affiliation{Department of Physics, University of Tokyo, Bunkyo, Tokyo 113-0033, Japan}

\author{Hiroyasu Matsuura}
\affiliation{Department of Physics, University of Tokyo, Bunkyo, Tokyo 113-0033, Japan}

\author{Masao Ogata}
\affiliation{Department of Physics, University of Tokyo, Bunkyo, Tokyo 113-0033, Japan}
\affiliation{Trans-scale Quantum Science Institute, University of Tokyo, Bunkyo, Tokyo 113-0033, Japan}

\begin{abstract}
    The spectral conductivity, i.e., the electrical conductivity as a function of the Fermi energy, is a cornerstone in determining the thermoelectric transport properties of electrons. However, the spectral conductivity depends on sample-specific properties such as carrier concentrations, vacancies, charge impurities, chemical compositions, and material microstructures, making it difficult to relate the experimental result with the theoretical prediction directly. Here, we propose a data-driven approach based on machine learning to reconstruct the spectral conductivity and chemical potential from the thermoelectric transport data. Using this machine learning method, we first demonstrate that the spectral conductivity and temperature-dependent chemical potentials can be recovered within a simple toy model. In a second step, we apply our method to experimental data in doped one-dimensional telluride Ta$_4$SiTe$_4$~[T. Inohara, \textit{et al.}, Appl. Phys. Lett. \textbf{110}, 183901 (2017)] to reconstruct the spectral conductivity and chemical potential for each sample. Furthermore, the thermal conductivity of electrons and the maximal figure of merit $ZT$ are estimated from the reconstructed spectral conductivity, which provides accurate estimates beyond the Wiedemann-Franz law. Our study clarifies the connection between the thermoelectric transport properties and the low-energy electronic states of real materials, and establishes a promising route to incorporate experimental data into traditional theory-driven workflows.
\end{abstract}

\maketitle

\section{Introduction}


Thermoelectric materials promise to be a future renewable energy source, where thermal energy is directly converted into electrical energy~\cite{yangThermoelectricMaterialsSpace2006,snyderComplexThermoelectricMaterials2008, zhangThermoelectricMaterialsEnergy2015a,ZhuCompromise2017,koumotoThermoelectric2013}. Despite the vast amount of studies to find a high-performance thermoelectric material~\cite{gaultoisDataDrivenReviewThermoelectric2013}, it is still challenging to optimize the dimensionless figure of merit $ZT$ owing to its dependence on conflicting material properties~\cite{snyderComplexThermoelectricMaterials2008}. One of the key ideas for suitable thermoelectric materials is to realize a ``phonon-glass electron-crystal"~\cite{snyderComplexThermoelectricMaterials2008}, thus requiring both phonon- and band-engineering.

\begin{figure}
    \centering
    \includegraphics[width=\columnwidth]{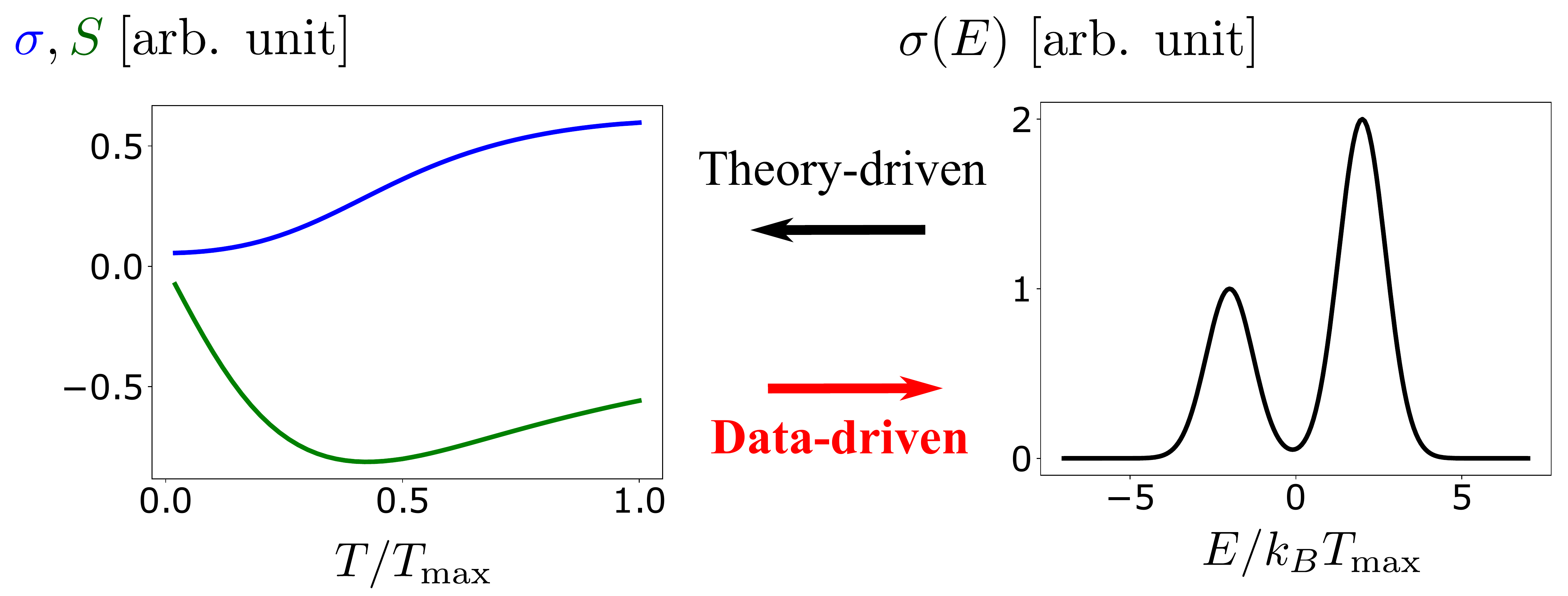}
    \caption{\textbf{Data-driven approach to reconstruct the spectral conductivity.} In a theory-driven approach, the electric conductivity $\sigma(T)$ (blue) and Seebeck coefficient $S(T)$ (green) are computed as a function of temperature from the spectral conductivity $\sigma(E)$. The spectral conductivity~$\sigma(E)$ can be obtained from the electronic band structure and phenomenological relaxation time. In our data-driven approach, we solve the inverse problem to deduce $\sigma(E)$ from the experimental data of $\sigma(T)$ and $S(T)$.
     }
    \label{fig: schematic}
\end{figure}

Theory-driven approaches to accelerate the search for candidate thermoelectric materials have attracted much attention recently~\cite{goraiComputationallyGuidedDiscovery2017}. The workflow of such a theory-driven approach is sketched in Fig.~\ref{fig: schematic}, where the electrical conductivity~$\sigma(T)$ and the Seebeck coefficient~$S(T)$ are computed from the spectral conductivity using linear response theory~\cite{sommerfeld2013elektronentheorie, mott1958theory, wilson1932theory, jonsonElectronphononContributionThermopower1990,kontaniGeneralFormulaThermoelectric2003,ogataRangeValiditySommerfeld2019}. Approximating the spectral conductivity~$\sigma(E,T)$ by the zero-temperature spectral conductivity~$\sigma(E)$, we can compute~$\sigma(E)$ and thermoelectric response functions from the electronic band structure~\cite{madsenBoltzTraPCodeCalculating2006,madsenBoltzTraP2ProgramInterpolating2018}. Hence, such a high-throughput approach based on first-principles density functional theory provides an efficient way to find new classes of high-performance thermoelectric materials~\cite{madsenAutomatedSearchNew2006, opahleHighThroughputDensity2012,beraIntegratedComputationalMaterials2014,zhuComputationalExperimentalInvestigation2015,bhattacharyaHighthroughputExplorationAlloying2015,matsumotoTwoPressureinducedSuperconducting2018}. However, this approach has a limitation because the thermoelectric coefficients depend on sample-dependent properties such as material microstructures, chemical compositions, vacancies, charge impurities, and carrier concentrations. While several approaches have been proposed to account for the charge impurities, such as the KKR-CPA-LDA method~\cite{akaiElectronicStructureNi1982,akaiFastKorringaKohnRostokerCoherent1989,schroterFirstprinciplesInvestigationsAtomic1995}, it is still difficult to predict $\sigma(T)$ and $S(T)$ accurately.

Alternatively, material informatics provides new data-driven approaches to reveal complex relations among material properties~\cite{rajanMaterialsInformatics2005,doi:10.1146/annurev-matsci-070214-021132,wanMaterialsDiscoveryProperties2019}. The data can be generated by experiments, high-throughput simulations, or combinations thereof. In addition, there are valuable repositories for material science communities, including Materials Projects~\cite{jainCommentaryMaterialsProject2013}, Aflow~\cite{curtaroloAFLOWAutomaticFramework2012}, MDF~\cite{blaiszikMaterialsDataFacility2016,blaiszikDataEcosystemSupport2019}, and QQMD~\cite{saalMaterialsDesignDiscovery2013,kirklinOpenQuantumMaterials2015}. By applying machine learning to the data obtained via ab-initio methods, several studies have successfully demonstrated the screening of materials for desired properties such as the low thermal conductivity~\cite{carreteFindingUnprecedentedlyLowThermalConductivity2014, sekoPredictionLowThermalConductivityCompounds2015} and narrow-gap band structures~\cite{matsumotoTwoPressureinducedSuperconducting2018}. In addition, the designs of nanostructures optimized for phonon transport were determined by combining Green function methods and Bayesian optimization~\cite{juDesigningNanostructuresPhonon2017}. Machine learning methods can also be used to accelerate computationally expensive calculations~\cite{yangMachineLearningArtificial2018}. However, similar to the theory-driven methods, these data-driven methods often ignore differences among samples~\cite{wanMaterialsDiscoveryProperties2019}, which makes it challenging to directly relate experimental results of $\sigma(T)$ and $S(T)$ with computational predictions.
Therefore, a method to incorporate the sample-dependent properties in theory- and data-driven approaches is needed.

Here, we propose a data-driven approach based on neural networks to reconstruct the spectral conductivity~$\sigma(E)$ and chemical potential $\mu(T)$ from experimental data of the electrical conductivity~$\sigma(T)$ and the Seebeck coefficient~$S(T)$. As shown in Fig.~\ref{fig: schematic}, our approach solves the problem inversely compared to the theory-driven approach. Although a similar idea was explored before~\cite{kurikideeplearning2019}, the key advantage in our method is that we can fully incorporate the sample dependence of the experimental data in $\sigma(E)$ and $\mu(T)$, giving a deeper insight into promising thermoelectric materials. Furthermore, $\sigma(E)$ reflects both the bulk band structure and electron scattering in materials. Thus, the reconstruction of $\sigma(E)$ is intriguing even beyond thermoelectric materials.

The structure of this paper is as follows. In Sec.~\ref{sec: model}, linear response theory for thermoelectric transport is briefly introduced. We define our data-driven approach to reconstruct $\sigma(E)$ and $\mu(T)$ from experimental data. This method is verified using test data generated from a toy model, where the spectral conductivity and chemical potentials of the toy model are successfully reproduced. In Sec.~\ref{section: result}, we apply our method to the experimentally acquired data of doped one-dimensional telluride Ta$_4$SiTe$_4$~\cite{inoharaLargeThermoelectricPower2017}. Using the reconstructed spectral conductivity and chemical potential, the maximal figure of merit $ZT$ is estimated for various doping concentrations. In Sec.~\ref{sec: conc}, we discuss the limitations of our method and provide a summary of this work.

\section{Theoretical framework and model}
\label{sec: model}
\subsection{Linear response theory for thermoelectric transport}
\label{sec: linear_response}
Within the framework of linear response theory, thermoelectric transport is described as a response to electric fields and thermal gradients~\cite{kuboStatisticalMechanicalTheoryIrreversible1957, luttingerTheoryThermalTransport1964,smrckaTransportCoefficientsStrong1977,cooperThermoelectricResponseInteracting1997}.
Electric currents and heat currents are given in terms of the linear response coefficients as~\cite{mahanManyParticlePhysics2000}
\begin{eqnarray}
\textbf{j}&=&L_{11}\textbf{E}+L_{12}\Big(-\frac{\boldsymbol{\nabla} T}{T}\Big),\\
\textbf{j}_Q&=&L_{21}\textbf{E}+L_{22}\Big(-\frac{\boldsymbol{\nabla} T}{T}\Big),
\end{eqnarray}
where $\textbf{j}$, $\textbf{j}_Q$, $\textbf{E}$, and $\boldsymbol{\nabla} T$ are electric current density, heat current density, electric field, and temperature gradient. While the electrical conductivity is given by \begin{equation}
    \sigma=L_{11},
    \label{eq: cond_L11}
\end{equation}
the Seebeck coefficient is defined as the voltage induced by a temperature gradient when $\textbf{j}=0$ and given by
\begin{equation}
    S=-\frac{\Delta V}{\Delta T}=\frac{L_{12}}{TL_{11}}.
    \label{eq: S_L12}
\end{equation}
Another important thermoelectric transport property is the thermal conductivity $\kappa$, which is defined from the thermal current densities subjected to the temperature gradient when $\textbf{j}=0$. From these thermoelectric properties, a dimensionless figure of merit is defined for thermoelectric materials as $ZT=S^2\sigma T/(\kappa_\textrm{el}+\kappa_\textrm{lat})$, where $\kappa_\textrm{el}$ and $\kappa_\textrm{lat}$ represent the contribution from electrons and phonons, respectively. In this paper, we will concentrate on the electronic contribution and neglect the phononic contribution.

In linear response theory, thermoelectric coefficients are given by the following Sommerfeld–Bethe relation~\cite{sommerfeld2013elektronentheorie, mott1958theory, wilson1932theory, jonsonElectronphononContributionThermopower1990,kontaniGeneralFormulaThermoelectric2003,ogataRangeValiditySommerfeld2019}:
\begin{subequations}
\begin{eqnarray}
L_{11}&=&\int^{\infty}_{-\infty}dE \left(-\frac{\partial f(E,T)}{\partial E}\right)\sigma(E,T),\label{L11}\\
L_{12}&=&-\frac{1}{e}\int^{\infty}_{-\infty}dE \left(-\frac{\partial f(E,T)}{\partial E}\right)(E-\mu)\sigma(E,T)\label{L12},\quad
\end{eqnarray}
\label{eq: SB_relation1}
\end{subequations}
where $\sigma(E,T)$ is the spectral conductivity, $e>0$ is the elementary electric charge, $\mu$ is the chemical potential, and $f(E,T)=1/(\exp [(E-\mu)/k_BT]+1)$ is the Fermi-Dirac distribution function. From Onsager’s reciprocal theorem, we have $L_{21}=L_{12}$~\cite{PhysRev.37.405, PhysRev.38.2265}. Note that the Sommerfeld–Bethe relation does not always hold. For example, the electron-phonon~\cite{jonsonElectronphononContributionThermopower1990, matsuuraEffectPhononDrag2019} and the electron-electron coupling~\cite{kontaniGeneralFormulaThermoelectric2003, ogataRangeValiditySommerfeld2019} can lead to anomalous terms. In this paper, we assume that the anomalous terms beyond the Sommerfeld–Bethe relation are negligible. Furthermore, the temperature dependence of the spectral conductivity is not significant in general as long as there is no phase transition. 
Hence, the spectral conductivity is assumed to be independent of temperatures in the following, replacing $\sigma(E,T)$ with $\sigma(E)$ as in Refs.~\cite{madsenBoltzTraPCodeCalculating2006,madsenBoltzTraP2ProgramInterpolating2018}.

From Eqs.~\eqref{L11} and \eqref{L12}, in the relaxation time approximation of Boltzmann's theory, the thermoelectric properties are determined by the spectral conductivity $\sigma(E)$, which is given by $\sigma(E)=e^2 \tau(E) v^2(E)D(E)$ in isotropic systems with the relaxation time $\tau(E)$, group velocity $v(E)$, and density of states $D(E)$~\cite{madsenBoltzTraPCodeCalculating2006, madsenBoltzTraP2ProgramInterpolating2018}. Typically, the relaxation time depends on impurity potentials of each experimental sample, making it challenging to determine $\tau(E)$ from an ab-initio approach~\cite{goraiComputationallyGuidedDiscovery2017}. 

\subsection{Neural network model}
\label{section: NN model}

Motivated by the limitation of ab-initio methods, we consider a data-driven, inverse approach to deduce the full energy dependence of $\sigma(E)$ from experimental data. Assuming that the temperature dependence of the chemical potential is negligible, it is straightforward to show that there is a unique solution $\sigma(E)$ for a given set of $L_{11}(T)$ and $L_{12}(T)$ from Eqs.~\eqref{L11} and~\eqref{L12}. It follows that there is a one-to-one correspondence between $L_{11(12)}(T)$ and $\sigma_\textrm{sym(anti)}(E)$, where $\sigma_\textrm{sym(anti)}(E)$ is the (anti)symmetric part of the spectral conductivity $\sigma(E)=\sigma_\textrm{sym}(E)+\sigma_\textrm{anti}(E)$. 
Therefore, we can rewrite Eqs.~\eqref{L11} and~\eqref{L12} as an integration from $E=0$ to $E=+\infty$. Upon discretization, we obtain
\begin{subequations}
\begin{equation}
    L_{11}(T_i)=\frac{2W_c}{N}\sum_{j=1}^{N} F(E_j,T_i)\sigma_\textrm{sym}(E_j),
    \label{eq: Mat_L11}
\end{equation}
\begin{equation}
    L_{12}(T_i)=\frac{2W_c}{N}\sum_{j=1}^N G(E_j,T_i)\sigma_\textrm{anti}(E_j),
    \label{eq: Mat_L12}
\end{equation}
\end{subequations}
where $F(E_j,T_i)=-\partial_E f(E_j,T_i)$ and $G(E_j,T_i)=(E_j-\mu)\partial_E f(E_j,T_i)/e$.
Here, we have truncated the integral at a high-energy cutoff $E_N=W_ck_BT_N$ with a dimensionless cutoff $W_c$ and introduced linearly spaced grids in temperature $T_i$ and energy $E_j=W_ck_BT_j$ for $i,j=1,2,\ldots,N$.

Naively, one could use the inverse of the matrices $F$ and $G$ in these expressions to obtain $\sigma_\textrm{sym}(E_j)$ and $\sigma_\textrm{anti}(E_j)$.
However, the matrices $F$ and $G$ have exponentially small determinants. Since the inverse matrices are proportional to the inverse of their determinants, any numerical solver for inverse matrices will fail to compute $F^{-1}$ and $G^{-1}$ due to the floating-point overflow~(see Appendix~\ref{section: determinant}).
Exponentially small determinants of $F$ and $G$ also result in strong amplification of noises in $\sigma(E_j)$. Thus, the inverse problems of Eqs.~\eqref{eq: Mat_L11} and~\eqref{eq: Mat_L12} are ill-conditioned, where the exact solution cannot be obtained in general. Nevertheless, it is still possible to infer an approximate solution of $\sigma(E_j)$ by applying the numerical methods developed for another ill-conditioned inverse problem: the analytic continuation problem of quantum-many body systems. They include the maximum entropy method~\cite{jarrellBayesianInferenceAnalytic1996,gunnarssonAnalyticalContinuationImaginary2010}, the stochastic method~\cite{sandvikStochasticMethodAnalytic1998}, and sparse modeling approaches~\cite{otsukiSparseModelingApproach2017, otsukiSparseModelingQuantum2020}. 
Recently, a projected regression method based on supervised learning was used to construct a solution from a large database of input-output pairs, taking advantage of fast forward calculations~\cite{arsenaultProjectedRegressionMethod2017}. This machine-learning method performed better than an optimized maximum entropy implementation.

\begin{figure}
    \centering
    \includegraphics[width=\columnwidth]{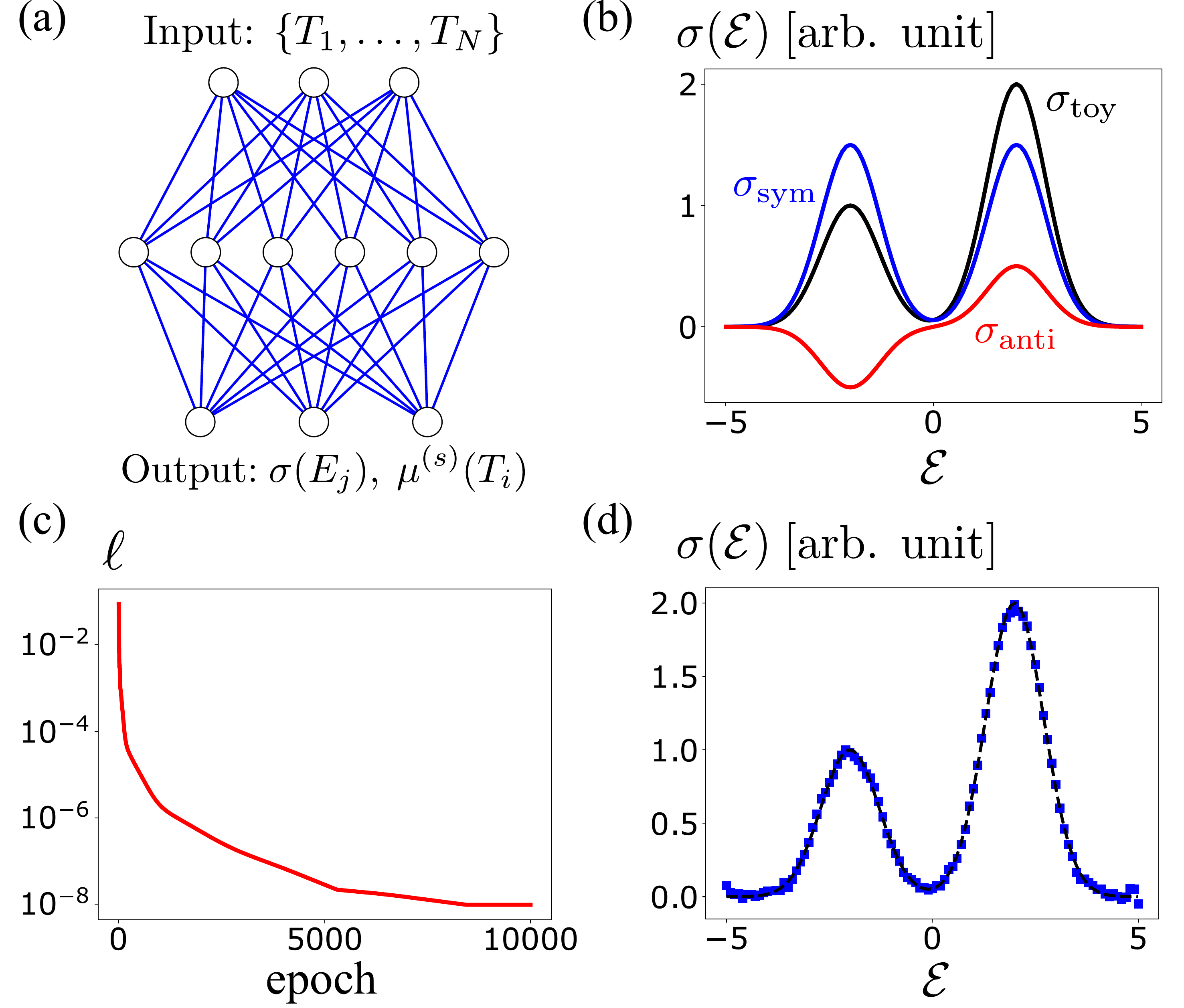}
    \caption{\textbf{A neural network can reconstruct the spectral conductivity.} (a) Schematic representation of the fully connected neural network (NN) with three layers~ (created using Ref.~\cite{lenailNNSVGPublicationReadyNeural2019}). 
    The NN takes a list of temperatures $\{T_1,\ldots,T_N\}$ as an input and returns the spectral conductivity $\sigma(E_j)$ and the chemical potential $\mu^{(s)}(T_i)$ for each sample.
    (b) The toy model $\sigma_\textrm{toy}(\CalE)$ (black) decomposed into a symmetric part~(blue), $\sigma_\textrm{sym}(\CalE)=\sigma_\textrm{sym}(-\CalE)$, and an antisymmetric part~(red), $\sigma_\textrm{anti}(\CalE)=-\sigma_\textrm{anti}(-\CalE)$ with $\CalE=E/k_BT_N$. (c)~Loss function $\ell$ as a function of the number of epochs. 
    (d) The reconstructed spectral conductivity using the NN (blue) is plotted with $\sigma_\textrm{toy}(\CalE)$~(black dashed). The NN is implemented in PyTorch~\cite{NEURIPS2019_9015}.
    }
    \label{fig: toymodel_cons}
\end{figure}

In this work, we propose an efficient machine learning algorithm to infer $\sigma(E_j)$. 
Our method constitutes a data-driven interpolating model and employs a neural network~(NN) at its core~(see Appendix~\ref{section: details_NN} for details). We first emphasize that our method does not need any prior training on data generated from theoretical models in contrast to Ref.~\cite{arsenaultProjectedRegressionMethod2017}.
The NN is trained to reproduce a given dataset~$\{L_{11}(T_i),L_{22}(T_i)\}$ to infer the spectral conductivity~$\sigma(E_j)$ without using the large database of input-output pairs. This advantage simplifies the implementation of our method in experimental studies. It also implies that our method can be used without knowledge about the microscopic origins of thermoelectric transport and the low-energy electronic states.

The structure of the employed NNs is sketched in Fig.~\ref{fig: toymodel_cons}(a).
The NN takes a list of equidistantly spaced temperatures $\{T_1,\ldots,T_N\}$ with grid spacing $T_1$ as an input, and returns a list of the spectral conductivity $\{\sigma(E_j)\}$ as an output. 
The output is then used to compute $\{L^{(s)}_{11}(T_i), L^{(s)}_{12}(T_i)\}$ for $i=1,2,\ldots, N$ and $s=1,2,\ldots,N_s$. Here, $N$ and $N_s$ denote the total number of data points for each sample and the total number of samples with different chemical doping concentrations, respectively. 
Importantly, the temperature dependence of chemical potentials $\mu^{(s)}(T_i)$ can be introduced
for $N_s>1$ as discussed below.

The values of $L^{(s)}_{11}(T_i)$ and $L^{(s)}_{12}(T_i)$ are evaluated by
\begin{subequations}
\begin{align}
L^{(s)}_{11,\textrm{ML}}(\xi_i)&\approx \int^{W_c \xi_i+\alpha^{(s)}_i}_{-W_c\xi_i+\alpha^{(s)}_i} d\CalE \frac{\sigma(\CalE)\exp[(\CalE-\alpha^{(s)}_i)/\xi_i]}{\xi_i (\exp[(\CalE-\alpha^{(s)}_i)/\xi_i]+1)^2}  , \label{eq:L11a}\\
L^{(s)}_{12,\textrm{ML}}(\xi_i)&\approx -\frac{k_BT_N}{e}\int^{W_c \xi_i+\alpha^{(s)}_i}_{-W_c\xi_i+\alpha^{(s)}_i} d\CalE\, (\CalE-\alpha^{(s)}_i)\nonumber \\
&\times\frac{\sigma(\CalE)\exp[(\CalE-\alpha^{(s)}_i)/\xi_i]}{\xi_i (\exp[(\CalE-\alpha^{(s)}_i)/\xi_i]+1)^2},
\label{eq:L12a}
\end{align}
\end{subequations}
where $\xi_i=T_i/T_N$, $\alpha^{(s)}_i=\mu^{(s)}(T_i)/k_BT_N$, and $\CalE=E/k_BT_N$ are dimensionless. The numerical integration is performed by linearly interpolating $\sigma(\CalE_j)$ using the cutoff $W_c$. For smoothing $\sigma(\CalE_j)$, we define $\sigma(\CalE_j)=\sum^{10}_{k=1}o_{j+k}/10$ for $j=-N,-N+1,\ldots, N$ $(N_s=1)$ or  $j=-2N,-2N+1,\ldots, 2N$ $(N_s>1)$, with $o_j$ representing an output of the NN.
Since the lowest temperature $T_1$ in the input sets a lower bound for the resolution of $\sigma(\CalE_j)$, the spacing in energy is fixed as $\CalE_j-\CalE_{j-1}=W_c \xi_1$ with $\CalE_j=W_c\xi_j$ for all $j$. Thus, the NN computes $\sigma(\CalE)$ for $|\CalE|\le W_c~(N_s=1)$ or $|\CalE|\le 2W_c~(N_s>1)$. The loss function is defined as the mean-squared error loss between estimated values and experimental values of $L_{11}$ and $L_{12}$, given as
\begin{align}
\ell&=
\sum_{i,s} a_i\left[\left\{L_{11}^{(s)}(\xi_i)-L_{11,\textrm{ML}}^{(s)}(\xi_i)\right\}^2\right.
\nonumber\\
&
+\left.\left\{\overline{L}_{12}^{\,(s)}(\xi_i)-\overline{L}_{12,\textrm{ML}}^{\,(s)}(\xi_i)\right\}^2\right]+b\left(\sum_j\textrm{min}\left[\sigma(\CalE_j),0\right]\right)^2,\label{eq: loss}
\end{align}
where $a_i$ and $b$ are hyperparameters. Here, we introduced $\overline{L}_{12}=eL_{12}/(k_BT_N)$, such that $L_{11}$ and $\overline{L}_{12}$ have the same physical unit.
The third term with $b>0$ is added to ensure the positivity of the spectral conductivity for $N_s>1$. It is also possible to enforce positive outputs from NNs using, e.g.,  the rectified linear activation function. However, we find empirically that the loss function~$\ell$ cannot be minimized efficiently when positive outputs are enforced via the design of the NNs. For $N_s=1$, where a unique solution of Eqs.~\eqref{eq:L11a} and~\eqref{eq:L12a} for $\mu=0$ exists, we found that the additional positivity constraints harms the optimization. Therefore, we take $b=0$ in this case. The trained NN approximates the spectral conductivity on a grid as described above.

If only single sample data are available $(N_s=1)$, the input is not sufficient to determine the chemical potential. This follows from the one-to-one correspondence between $L_{11(12)}(T)$ and $\sigma_\textrm{sym(anti)}(E)$ as discussed above, implying that the degree of freedom of output becomes greater than the number of constraints from the input if the chemical potential is taken into account. Thus, we assume the chemical potential to be a constant~$(\mu=0)$. In contrast, the temperature dependence of the chemical potential $\mu^{(s)}(T)$ can be included and inferred for $N_s>1$. Assuming that the spectral conductivity is unchanged by a small chemical doping and thus the same for all samples, the chemical potential of each sample is estimated by
\begin{equation}
    \alpha^{(s)}_i=\frac{2W_c}{\pi}\arctan (c^{(s)}_0+c^{(s)}_1\xi_i^2),\label{eq: chemical_model}
\end{equation}
where $c^{(s)}_0$ and $c^{(s)}_1$ are determined by the NN with additional outputs. The coefficient of the linear term in $\xi_i$ is set to zero based on the Sommerfeld expansion. Since the amplitude of $\alpha^{(s)}_i$ is bounded as $|\alpha^{(s)}_i|<W_c$ by the arctan function, both upper and lower cutoffs for integrals in Eqs.~\eqref{eq:L11a} and~\eqref{eq:L12a} are taken within $|\CalE|<2W_c$ for $N_s>1$. This ensures the range of integrals to be within the bounds of $\CalE_j$. This approach is valid as long as a small amount of chemical doping is carried out, resulting in a slight change of the chemical potentials.

\begin{figure}[t]
    \centering
    \includegraphics[width=\columnwidth]{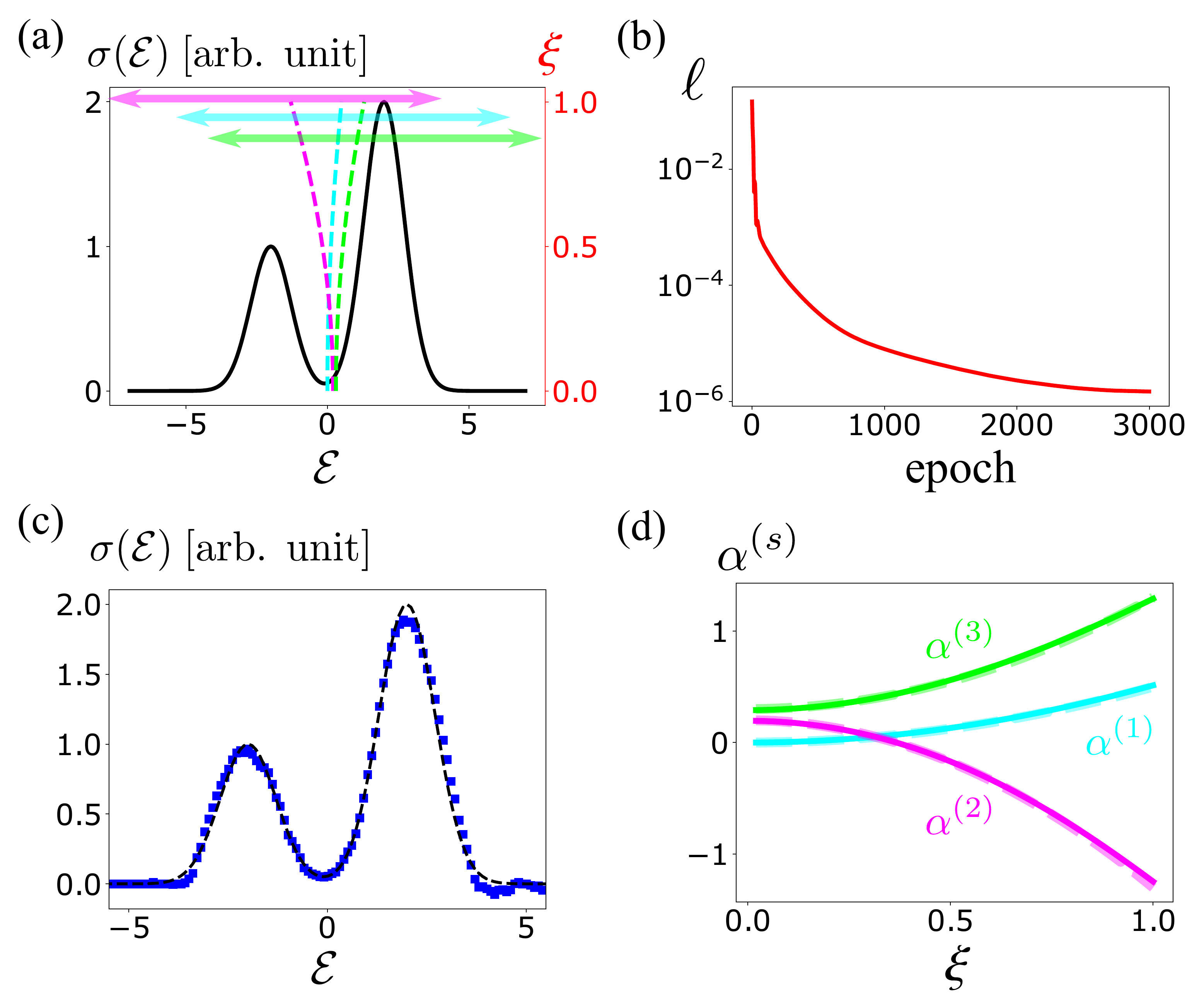}
    \caption{\textbf{Learning the temperature dependence of chemical potentials in the toy model.} (a) Toy model $\sigma_\textrm{toy}(\CalE)$ (black solid) with chemical potentials $\alpha_{\textrm{toy},i}^{(s)}$ for $s=1,2,3$, which are plotted as dashed cyan, dashed magenta, and dashed lime lines, respectively, against the dimensionless temperature $\xi_i=T_i/T_N$ on the right vertical axis. Each arrow indicates the range of integrals for evaluating $L_{11}^{(s)}$ and $L_{12}^{(s)}$ at $\xi_i=1$.
    (b) Loss function and (c) reconstructed spectral conductivity (blue) with $\sigma_\textrm{toy}(\CalE)$~(black dashed) using the NN. (d)~Temperature dependence of chemical potentials~$\alpha_i^{(s)}$~(solid) predicted by the NN compared with $\alpha_{\textrm{toy},i}^{(s)}$~(dashed). }
    \label{fig: toymodel_chem}
\end{figure}

\subsection{Performance test with toy model}
\label{section: toymodel}
Before applying our machine learning method to the experimental data, we investigate its ability to reconstruct the spectral conductivity and chemical potentials using a toy model. 
The toy model contains two Gaussian peaks with different amplitudes, defined as
\begin{equation}
    \sigma_\textrm{toy}(\CalE)=2\exp[-(\CalE-2)^2]+\exp[-(\CalE+2)^2] ,
    \label{eq: sigma_toy}
\end{equation}
and is displayed in Fig.~\ref{fig: toymodel_cons}(b).
Using this toy model, we generate training datasets with different chemical potentials and evaluate the validity of our approach. In this section, we fix $W_c=5$ and $a_i=1$. The size of training datasets is chosen to be $N=50$.

First, we consider $N_s=1$ with a constant chemical potential $(\mu=0)$. 
The spectral conductivity is decomposed into a symmetric part $\sigma_\textrm{sym}(\CalE)=\sigma_\textrm{sym}(-\CalE)$ and an antisymmetric part  $\sigma_\textrm{anti}(\CalE)=-\sigma_\textrm{anti}(-\CalE)$, indicated by blue and red solid lines in Fig.~\ref{fig: toymodel_cons}(b), respectively. As discussed in Sec.~\ref{section: NN model}, the symmetric and antisymmetric parts are determined from $L_{11}(\xi)$ and $L_{12}(\xi)$, respectively. 
Figure~\ref{fig: toymodel_cons}(c) shows that the loss function dropped immediately and converged around 10$^{-8}$ after 10000 epochs. Here, the initial learning rate $\lambda_0$ was set at $10^{-3}$ and gradually decreased as discussed in Appendix~\ref{section: details_NN}.
The reconstructed spectral conductivity shows a very good agreement with $\sigma_\textrm{toy}(\CalE)$, clearly reproducing smooth Gaussian peaks as shown in Fig.~\ref{fig: toymodel_cons}(d). Therefore, the NN can efficiently reconstruct the spectral conductivity assuming a constant chemical potential.

Second, we consider the temperature dependence of chemical potentials. Taking $N_s=3$, the chemical potential for each sample is chosen as $\alpha_{\textrm{toy},i}^{(1)}=0.5\xi_i^2$ (cyan), $\alpha_{\textrm{toy},i}^{(2)}=0.2-1.5\xi_i^2$ (magenta), and $\alpha_{\textrm{toy},i}^{(3)}=0.3+\xi_i^2$ (lime)~[see Fig.~\ref{fig: toymodel_chem}(a)]. For each chemical potential $\alpha_{\textrm{toy},i}^{(s)}$, we compute $L_{11}^{(s)}(\xi)$ and $L_{12}^{(s)}(\xi)$ using Eqs.~\eqref{eq:L11a} and~\eqref{eq:L12a} to prepare the training data. Compared to the previous example with $N_s=1$, the NN has more degrees of freedom. Hence, we introduce another constraint by adding a penalty for negative values of $\sigma(\CalE)$ with $b=1/(2N+1)$ in Eq.~\eqref{eq: loss}. 
As shown in Fig.~\ref{fig: toymodel_chem}(b), the loss function is saturated at a sufficiently small value~$\ell\approx 2\times 10^{-6}$ with $\lambda_0=10^{-3}$. Furthermore, Fig.~\ref{fig: toymodel_chem}(c) and (d) demonstrate a successful reconstruction of the spectral conductivity and chemical potentials, although there are small deviations in the chemical potentials near $\xi\sim 1$. To reconstruct the chemical potentials up to high temperatures, we need to further minimize the loss function. One way is to introduce a small correction in chemical potentials, as implemented in the following section.

\section{Analysis using experimental data}
\label{section: result}

For practical applications, we need thermoelectric transport data over a wide range of temperatures. As discussed in Sec.~\ref{section: NN model}, the lowest temperature of data determines the resolution of spectral conductivity, while the highest temperature determines the energy range where $\sigma(E)$ can be obtained. In addition, a systematic study of various doping concentrations is desirable to account for the temperature dependence of chemical potentials. Based on these requirements, we have chosen the thermoelectric transport data of doped one-dimensional telluride Ta$_4$SiTe$_4$ for this work, where a large value of the Seebeck coefficient was observed~\cite{inoharaLargeThermoelectricPower2017}.

\begin{figure}
    \centering
    \includegraphics[width=\columnwidth]{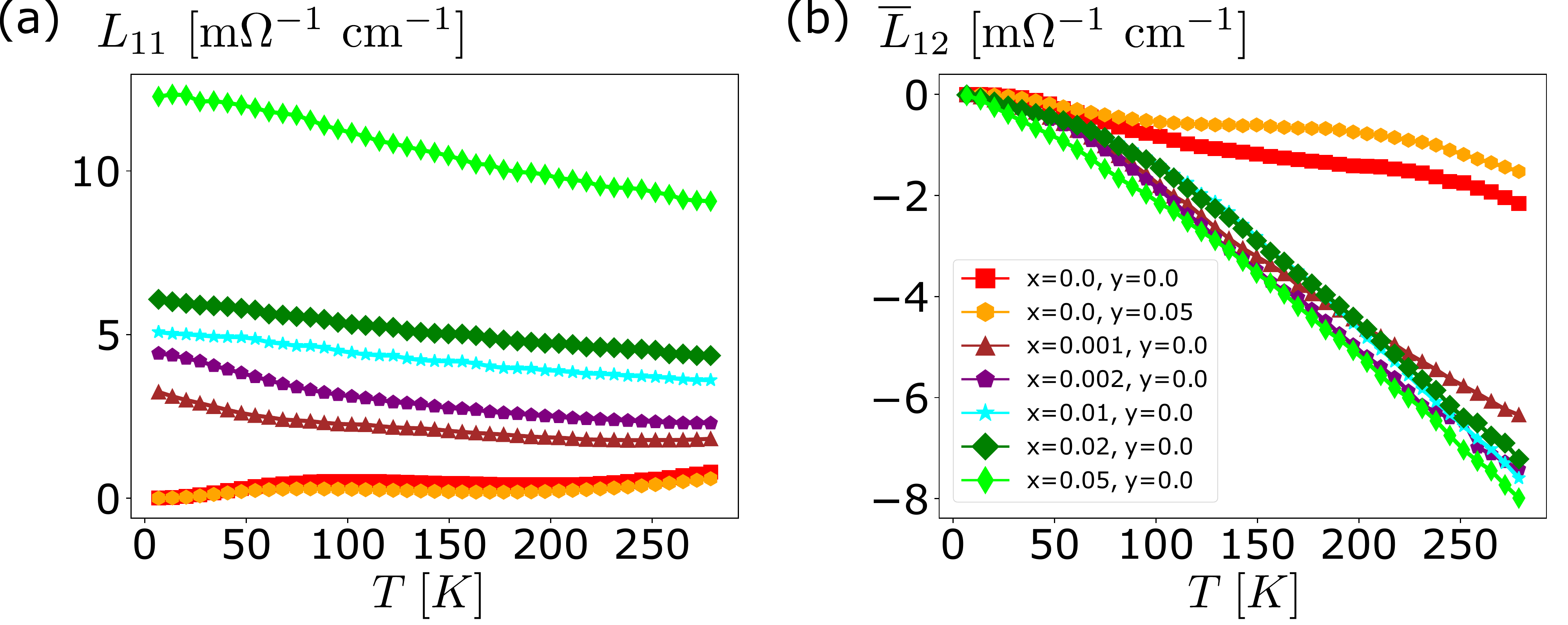}
    \caption{\textbf{Thermoelectric transport data of doped Ta$_4$SiTe$_4$, extracted from Ref.~\cite{inoharaLargeThermoelectricPower2017}.} (a,b) Temperature dependence of (a) $L_{11}$ and (b) $\overline{L}_{12}=eL_{12}/(k_BT_N)$, which is computed from the electrical resistivity and Seebeck coefficient of doped Ta$_4$SiTe$_4$ with $T_N=278.8$~K. In (b), the doping concentration of (Ta$_{1-x}$Mo$_x$)$_4$Si(Te$_{1-y}$Sb$_y$)$_4$ for each data is also shown.
    }
    \label{fig: exp_data}
\end{figure}

In order to generate a training dataset, we have extracted numerical data points from the figures in Ref.~\cite{inoharaLargeThermoelectricPower2017} using the method described in~\cite{Rohatgi2020}. The extracted experimental data are linearly interpolated to obtain the training dataset with $N=41$, $T_\textrm{1}=6.8$~K, and $T_N=278.8$~K. Figure~\ref{fig: exp_data} shows $L_{11}$ and $\overline{L}_{12}=eL_{12}/(k_BT_N)$ as a function of temperature, which is computed from the experimental data of the electrical conductivity and Seebeck coefficient according to Eqs.~\eqref{eq: cond_L11} and~\eqref{eq: S_L12}. Solid lines with different symbols represent various doping concentrations for (Ta$_{1-x}$Mo$_x$)$_4$Si(Te$_{1-y}$Sb$_y$)$_4$ with corresponding values of $x$ and $y$ given in Fig.~\ref{fig: exp_data}(b). Increasing the value of $x (y)$ corresponds to an increase of electron(hole)-doping. Without chemical doping, Ta$_4$SiTe$_4$ behaves as an insulator with a decrease in the electrical conductivity $L_{11}$ at low temperatures. This behavior is suppressed as it becomes metallic for larger values of $x$.

\begin{figure}[t]
    \centering
    \includegraphics[width=\columnwidth]{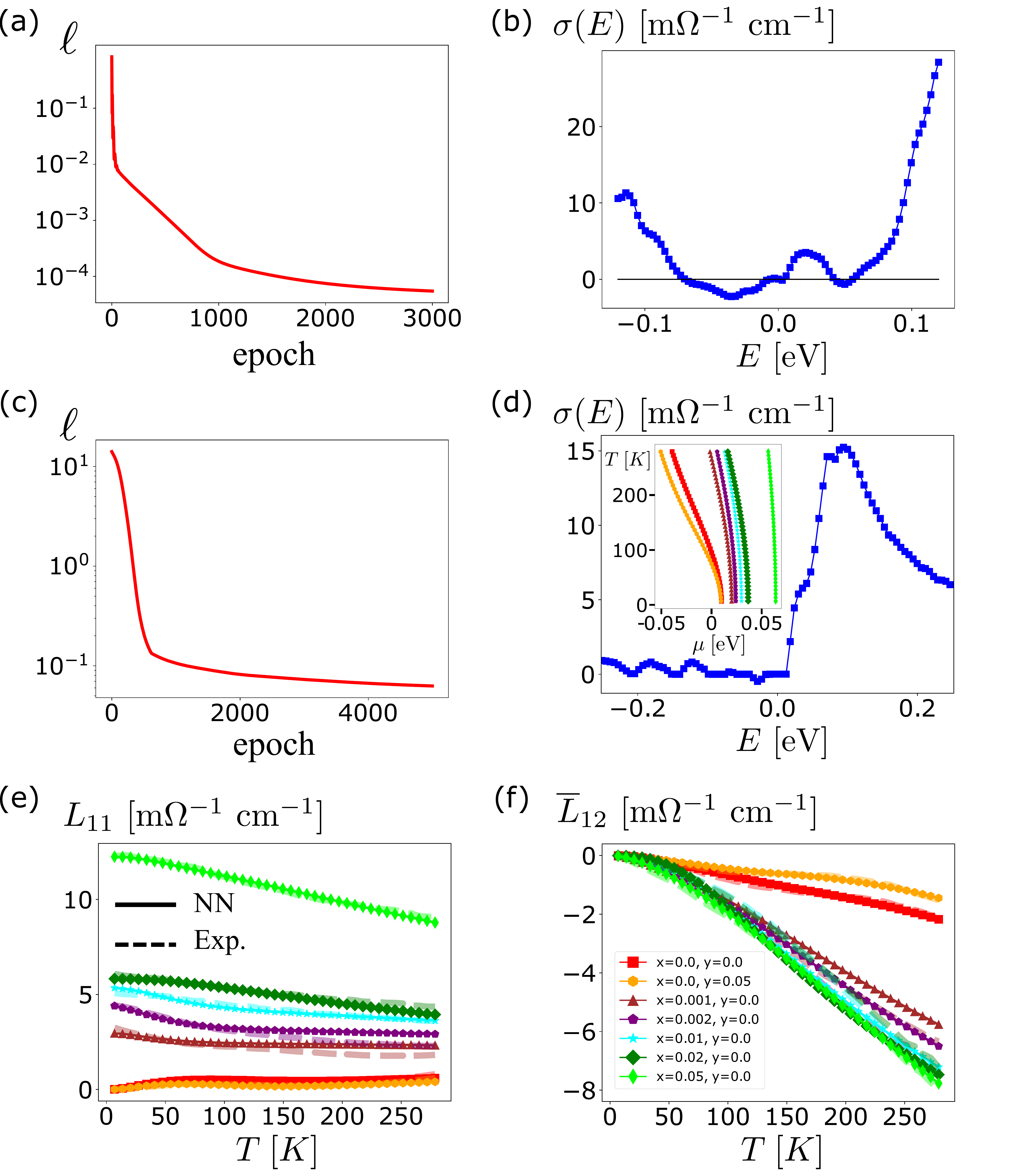}
    \caption{\textbf{Reconstructed spectral conductivity and chemical potentials of doped Ta$_4$SiTe$_4$.} (a)~Loss function and (b) reconstructed $\sigma(E)$ of Ta$_4$SiTe$_4$ using the NN. Here, we assume the chemical potential to be constant over temperature. (c) Loss function and (d) reconstructed $\sigma(E)$ (blue) of (Ta$_{1-x}$Mo$_x$)$_4$Si(Te$_{1-y}$Sb$_y$)$_4$ using the NN. In the inset of (d), symbols represent chemical potentials as a function of temperatures. 
    (e,f)~Temperature dependence of (e) $L_{11}$ and (f) $\overline{L}_{12}$ shown as solid lines with symbols, which are computed from $\sigma(E)$ and $\mu(T)$ in (d). The experimental data are shown as dashed lines. In (f), the doping concentration for each symbol is indicated for (d-f). }
    \label{fig: all_data_ML}
\end{figure}

\begin{figure*}
    \centering
    \includegraphics[width=180mm]{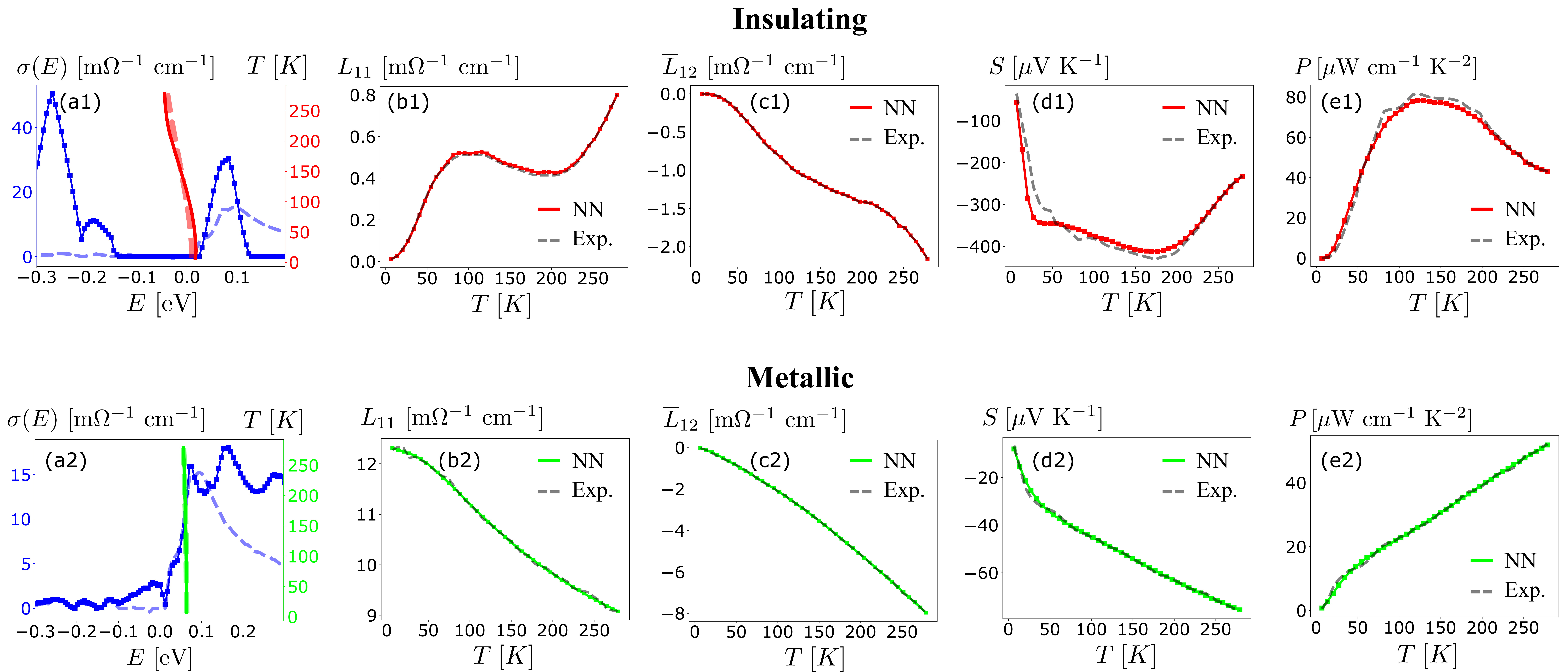}
    \caption{\textbf{Optimization of the spectral conductivity and chemical potential for doped Ta$_4$SiTe$_4$.}
    (a1,a2)~Reconstructed $\sigma^{(s)}(E)$ (solid blue with squares) of (Ta$_{1-x}$Mo$_x$)$_4$Si(Te$_{1-y}$Sb$_y$)$_4$ with $\mu^{(s)}(T)$ (solid red/lime) plotted on the right vertical axis. The initial condition is chosen as $\sigma_0(E)$ (dashed blue) and $\mu_0^{(s)}(T)$ (dashed red/lime) from Fig.~\ref{fig: all_data_ML}(d). The details of the computations are explained in the text. 
    (b1-e1,b2-e2)~Temperature dependence of (b1,b2)~$L_{11}$, (c1,c2)~$\overline{L}_{12}$, (d1,d2)~Seebeck coefficient~$S=L_{12}/(TL_{11})$, and (e1,e2)~power factor $P=\sigma S^2$, which are computed from $\sigma^{(s)}(E)$ and $\mu^{(s)}(T)$ in (a1) and (a2), respectively. In (b1-e1) and~(b2-e2), the results of the NN~(solid) are compared to the experimental data~(dashed).
    (a1-e1) are obtained for an insulating sample with $x=y=0$~(red) and (a2-e2) are obtained for a metallic sample with $x=0.05$ and $y=0$~(lime). 
    }
    \label{fig: sigE_both}
\end{figure*}

\subsection{Spectral conductivity with a constant chemical potential ($N_s=1$)}
Assuming that the chemical potential is constant over the considered range of temperatures, we study the spectral conductivity of undoped Ta$_4$SiTe$_4$ with $x=y=0$ (red lines in Fig.~\ref{fig: exp_data}). The loss function is minimized with $a_i=1$, $b=0$, $W_c=5$, and $\lambda_0=5\times10^{-3}$. As shown in Fig.~\ref{fig: all_data_ML}(a), the loss function is saturated below $10^{-4}$ after 3000 epochs, resulting in a good agreement with the experiment. The reconstructed spectral conductivity in Fig.~\ref{fig: all_data_ML}(b) exhibits a larger contribution from the conduction band than the valence band. In addition, the small peak inside the band gap at~$E\approx 0.02$~eV might indicate the formation of an impurity band due to vacancies and impurities. However, it contains unphysical negative contributions of $\sigma(E)$ at $E<0$ when $\sigma_\textrm{anti}(E)>\sigma_\textrm{sym}(E)$. 
These contributions imply that the experimental data cannot be described by a vanishing chemical potential. Thus, we introduce temperature-dependent chemical potentials.

\subsection{Spectral conductivity with temperature-dependent chemical potentials ($N_s=7$)}
\label{sec: sigmaE_N=7}
Next, the thermoelectric transport data of all doping concentrations in Fig.~\ref{fig: exp_data} is used for training of the NN with~$N_s=7$. The hyperparameters are chosen as $a_i=4/(4+\xi_i)$, $b=1$, $W_c=10$, and $\lambda_0=10^{-4}$, where the weight $a_i$ is increased at low temperatures to improve the precision of $\sigma(E)$ near the Fermi energy of each sample. After 5000 epochs, the loss function is saturated around $10^{-1}$ as shown in Fig.~\ref{fig: all_data_ML}(c). The obtained spectral conductivity and chemical potentials are shown in Fig.~\ref{fig: all_data_ML}(d) as a blue solid line and the corresponding symbols for each sample, respectively. We find that the Fermi energies move deep inside the bulk conduction band from the band edge as the value of $x$ is increased, indicating the transition from insulators to metals. This is qualitatively consistent with the experimental data. However, a comparison with the experimental data shows some errors in $L_{11}$ and $L_{12}$ at higher temperatures as shown in Fig.~\ref{fig: all_data_ML}(e) and (f).

\subsection{Optimized spectral conductivity with the temperature-dependent chemical potential ($N_s=1$)}
\label{sec: sigmaE_optimized}
For further improvement of the NN, we need to account for a change in the spectral conductivity of doped samples. While the electronic band structure is expected to remain almost unchanged with doping (rigid band approximation), the relaxation time $\tau(E)$ is generally modified upon doping.
However, simultaneous learning of the spectral conductivity and chemical potential leads to the problem of overfitting as discussed in Sec.~\ref{section: NN model}. In this work, we avoid this problem by imposing constraints on the chemical potentials, assuming that those obtained in Fig.~\ref{fig: all_data_ML}(d) are very close to the optimal solution.
Denoting the results obtained in Fig.~\ref{fig: all_data_ML}(d) as $\sigma_\textrm{0}(\CalE_j)$ and $\alpha^{(s)}_{0}$, we use the NN to estimate the sample-dependent spectral conductivity $\delta \sigma^{(s)}(\CalE_j)$. 
The spectral conductivity is given as $\sigma^{(s)}(\CalE_j)=\sigma_0(\CalE_j)+\delta \sigma^{(s)}(\CalE_j)$ with $\delta \sigma^{(s)}(\CalE_j)=\sum_{k=1}^{10}o_{j+k}/10$ for smoothing, where $o_i$ represents the $i$th component of output from the NN.
In addition, we allow a small correction in chemical potential $\delta \alpha^{(s)}$ to improve the accuracy of the NN, where the chemical potential is defined as $\alpha^{(s)}_i=\alpha_{0,i}^{(s)}+\delta \alpha^{(s)}_{i}$ with $\delta  \alpha^{(s)}_{i}=2\Delta\arctan (o_{4N+10+i})/\pi$. The small parameter $\Delta$ is gradually increased until we obtain a good agreement between the NN and experimental data. 

Taking $a_i=1$, $b=1$, and $W_c=10$, the NN is optimized for an insulating sample ($x=y=0$) with $\lambda_0=10^{-3}$ and a metallic sample ($x=0.05,\, y=0$) with $\lambda_0=10^{-4}$. 
Figure~\ref{fig: sigE_both}(a1) and (a2) show the results obtained after 5000 epochs, where the values of $\Delta$ are taken as $\Delta=0.04W_c$ and $\Delta=0$, respectively. 
Remarkably, we find that the NN reproduces the temperature dependence of chemical potentials in insulators without any prior knowledge~[see Fig.~\ref{fig: sigE_both}(a1)], where the chemical potential is shifted from the band edge to the middle of band gap as the temperature increases.
For the metallic sample with $x=0.05$ and $y=0$~[see Fig.~\ref{fig: sigE_both}(a2)], the chemical potential is fitted well by a quadratic term in temperatures. For both cases, the difference between $\mu_0^{(s)}$ and $\mu^{(s)}$ is not significant, as shown by red (lime) dashed and solid lines for~$x=y=0$~($x=0.05,\, y=0$), respectively. This implies that the result obtained in Fig.~\ref{fig: all_data_ML}(d) provides a good initial condition and thus prevents the overfitting problem.

The sample-dependent correction in the spectral conductivity is more prominent than in chemical potentials. Crucially, it provides deep insights of the low-energy electronic band structure. At $x=y=0$~[see Fig.~\ref{fig: sigE_both}(a1)], we find that there is a large contribution from the conduction band in the spectral conductivity. In addition, a small peak at $E=-0.2$~eV shows contributions from the valence band. The band gap is estimated as $0.15$~eV, which is consistent with the theoretically predicted value $\Delta_g\sim  0.1$~eV~\cite{inoharaLargeThermoelectricPower2017}. Although the impurity band is absent in Fig.~\ref{fig: sigE_both}(a1) with $\mu=\mu(T)$ in contrast to Fig.~\ref{fig: all_data_ML}(b) with $\mu=0$, the Fermi energy is placed at the band edge of the conduction band at $T=0$~K, implying a little doping or presence of a small impurity band in the experimental sample.
At $x=0.05$ and $y=0$~[see Fig.~\ref{fig: sigE_both}(a2)], the conduction-band contributions are dominant. In addition, we find several peaks of impurity bands at $E<0$.
The sample-dependent optimization greatly improves the agreement between the NN and experimental data in $L_{11}$, $\overline{L}_{12}$, the Seebeck coefficient~$S$, and power factor $P$, as shown in Fig.~\ref{fig: sigE_both}(b1-e1) and (b2-e2). Remarkably, even a complex temperature dependence of $P$ was successfully reproduced in Fig.~\ref{fig: sigE_both}(e1). However, the NN overestimates the Seebeck coefficient $S=L_{12}/(TL_{11})$ of the insulating sample at low temperatures as shown in Fig.~\ref{fig: sigE_both}(d1). This is because the values of both $L_{11}$ and $L_{12}$ approach zero as the temperature decreases~[see Fig.~\ref{fig: sigE_both}(b1) and~(c1)], rendering the computation of $S$ sensitive to small numerical errors in $L_{11}$ and $L_{12}$.

\begin{figure}
    \centering
    \includegraphics[width=\columnwidth]{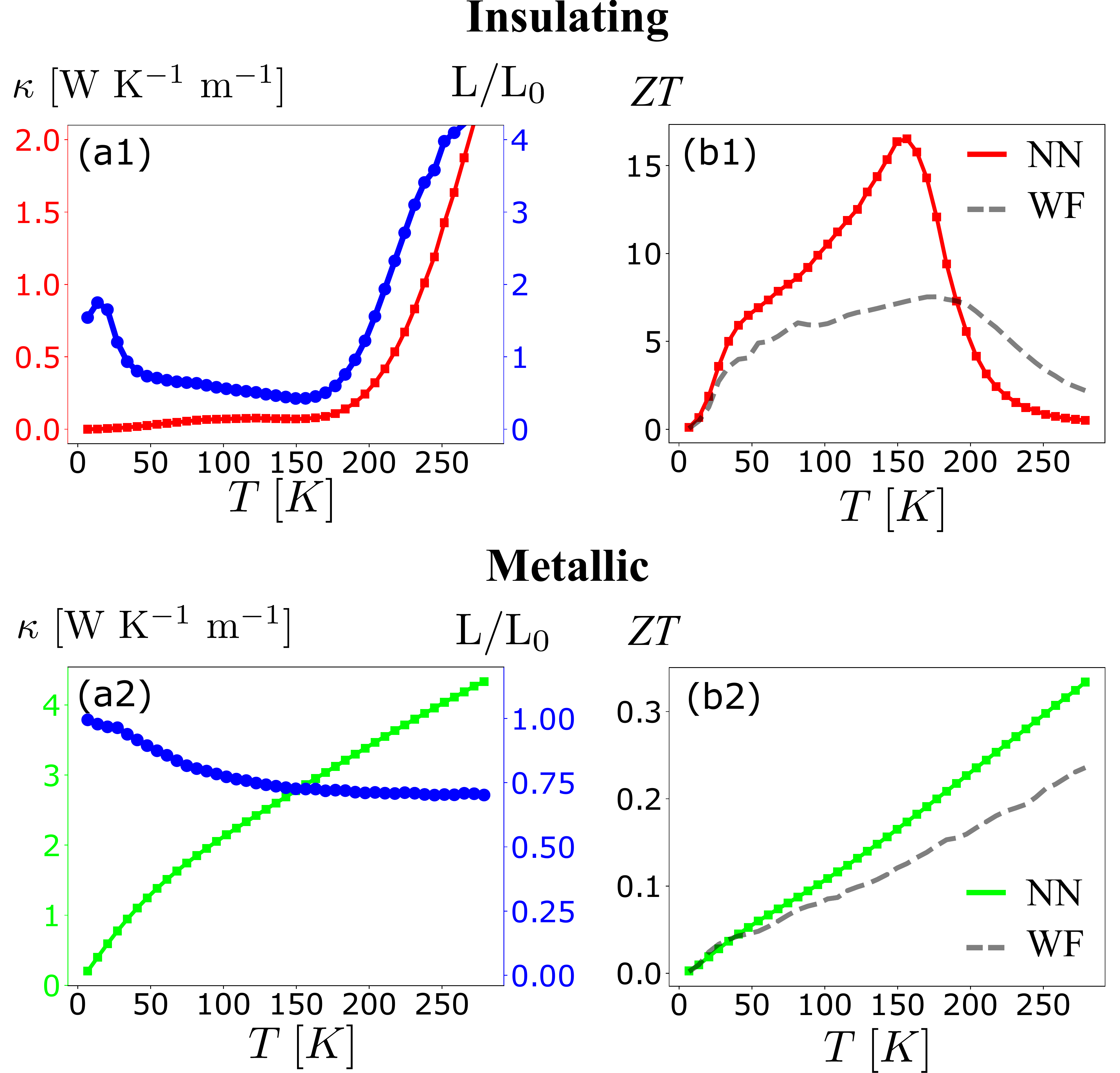}
    \caption{\textbf{Prediction of the maximal $ZT$ factor in doped Ta$_4$SiTe$_4$.}
    (a1,a2)~Electronic thermal conductivity $\kappa_\textrm{el}$ predicted by the NN~(solid with squares). On the right vertical axis, the ratio between the Lorenz number $L=\kappa_\textrm{el}/(TL_{11} )$ and universal constant $L_0=\pi^2k_B^2/(3e^2)$ is plotted~(solid with circles).
    (b1,b2)~Maximal figure of merit $ZT=S^2\sigma T/\kappa_\textrm{el}$ predicted by the NN~(solid with squares) in comparison with the Wiedemann-Franz law~(dashed).
    The results in (a1-b1) and (a2-b2) are computed from $\sigma^{(s)}(E)$ and $\mu^{(s)}(T)$ in (a1) and (a2) of Fig.~\eqref{fig: sigE_both}, respectively.
    }
    \label{fig: ZT_both}
\end{figure}

We can also predict the maximal $ZT$ value by computing the electronic thermal conductivity $\kappa_\textrm{el}$, which is given by
\begin{align}
    \kappa_\textrm{el}&=\frac{1}{T}\Big(L_{22}-\frac{L_{12}^2}{L_{11}}\Big),
    \label{eq: kappa}
\end{align}
with 
\begin{align}
    L_{22}&=\frac{1}{e^2}\int^{\infty}_{-\infty}dE \left(-\frac{\partial f(E,T)}{\partial E}\right)(E-\mu)^2\sigma(E)\label{L22}.
\end{align}
Since it is not possible to directly measure $\kappa_\textrm{el}$ in experiments, it is commonly estimated by the the Wiedemann–Franz (WF) law. According to the WF law, the thermal conductivity is related to the electrical conductivity by $\kappa_\textrm{el}=L_0TL_{11}$, where $L_0=\pi^2k_B^2/(3e^2)$ is the universal constant. Since it is based on the Sommerfeld expansion, the WF law usually provides good estimates for metals at low temperatures. However, it is known to break down for semiconductors and insulators~\cite{tobererAdvancesThermalConductivity2012}, which makes it challenging to estimate the phononic thermal conductivity~$\kappa_\textrm{lat}$. Remarkably, our data-driven approach offers more accurate estimates using the reconstructed spectral conductivity and chemical potential as shown below.

The breakdown of the WF law is characterized by the ratio between the Lorenz number~$L=\kappa_\textrm{el}/(TL_{11})$ and $L_0$.
In Fig.~\ref{fig: ZT_both}(a1) and (a2), we plot  $\kappa_\textrm{el}$ predicted by the NN~(red/lime solid with squares) with $L/L_0$ on the right vertical axis~(blue solid). As expected for insulators, Fig.~\ref{fig: ZT_both}(a1) shows a large deviation in $L/L_0$ from unity at $x=y=0$. Initially, $L/L_0$ decreases with temperatures until it reaches the minimum value $L/L_0\approx 0.42$ at $T\sim 150$~K.
When $L/L_0$ reaches the minimum, we find that $\kappa_\textrm{el}$ becomes almost flat between 100~K and 170~K, which is caused by the shift of the chemical potential into the middle of the band gap~[see Fig.~\ref{fig: sigE_both}(a1)]. Within this temperature range, $L_{11}$ decreases slightly while $\overline{L}_{12}$ is linear in $T$~[see Fig.~\ref{fig: sigE_both}(b1) and~(c1)]. The decrease in $L_{11}$ enhances the second term of Eq.~\eqref{eq: kappa}, cancels out higher order contributions in $T$ and leads to the flatness of $\kappa_\textrm{el}$.
At higher temperatures, $L_{11}$ starts to increase with $T$ due to the thermal excitation of electrons from the conduction and valance bands.
As a result, $\kappa_\textrm{el}$ and $L/L_0$ increase dramatically and becomes greater than unity at $T\sim 190$~K. 
For the metallic case at $x=0.05$ and $y=0$~[see Fig.~\ref{fig: ZT_both}(a2)], the WF law is in a good agreement with the NN at low temperatures. As the temperature increases, $L/L_0$ gradually decreases with $L/L_0\sim 0.75$ at $T> 150$~K.

Finally, the maximal $ZT$ factor is estimated from $ZT=S^2\sigma T/\kappa_\textrm{el}=S^2/L$. We should note that the lattice contribution $\kappa_\textrm{lat}$ is neglected here~(see Sec.~\ref{sec: linear_response}). Thus, the actual value of $ZT$ can be much smaller depending on $\kappa_\textrm{lat}$.
For the insulating case with $x=y=0$~[see Fig.~\ref{fig: ZT_both}(b1)], the NN predicts a large value of $ZT$ factor with $ZT>10$. Furthermore, it exhibits a sharp peak at $T\approx 150$~K, which arises from the flattening of $\kappa_\textrm{el}$. Since $L/L_0$ is smaller than unity at this temperature, the prediction by the WF law is smaller than the NN and does not exhibit a peak. Above 170~K, the  $ZT$ factor drops sharply due to a rapid increase in $\kappa_\textrm{el}$.
At $x=0.05$ and $y=0$~[see Fig.~\ref{fig: ZT_both}(b2)], the $ZT$ factor is approximately the same for the NN and WF law as expected for metals, showing a linear increase with temperatures.

\begin{figure}[t]
    \centering
    \includegraphics[width=\columnwidth]{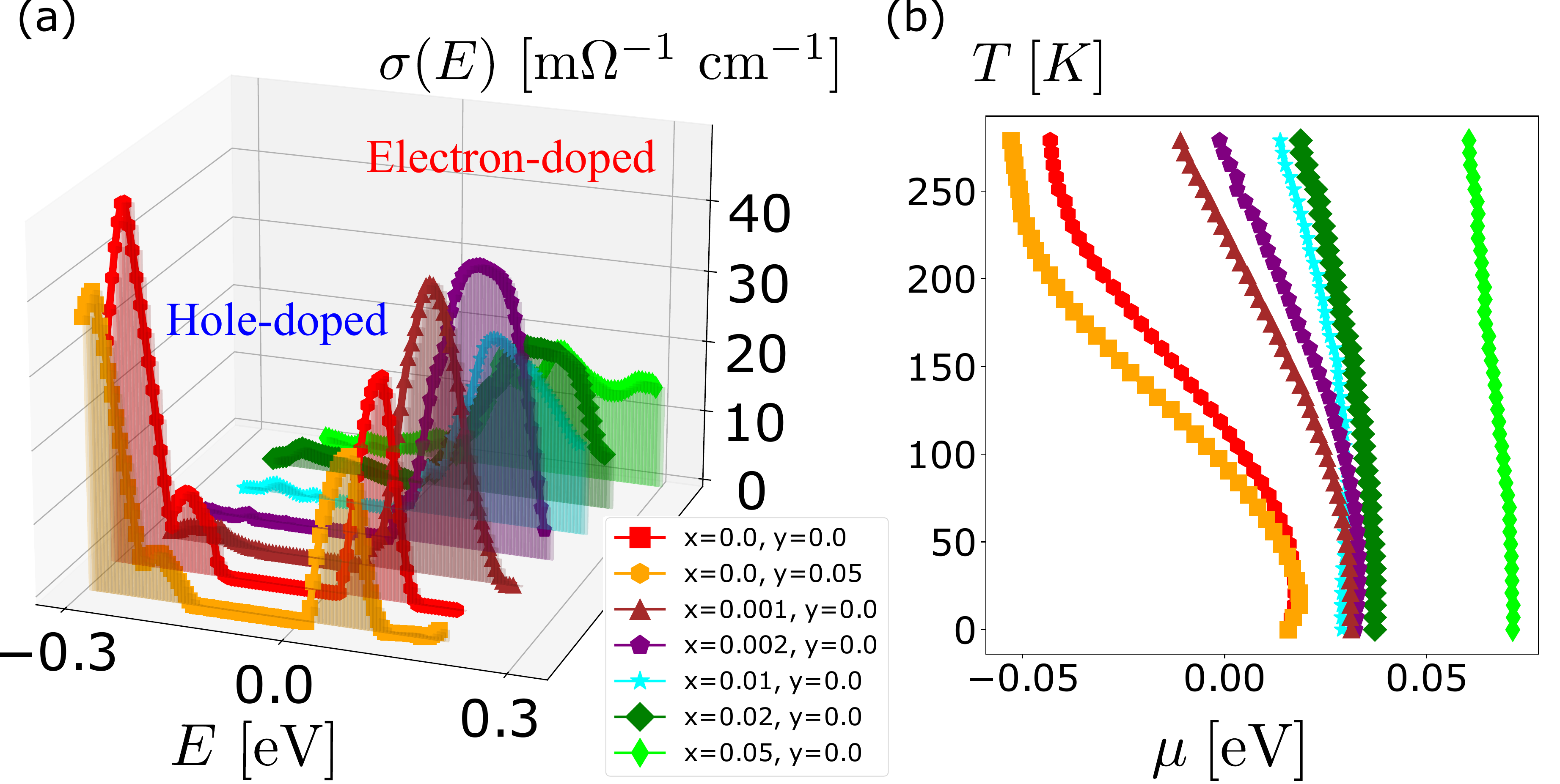}
    \caption{\textbf{Chemical doping dependence of spectral conductivity and chemical potential in Ta$_4$SiTe$_4$.} (a)~Spectral conductivity $\sigma^{s}(E)$ and (b)~chemical potential $\mu^{s}(T)$ is plotted for various doping concentrations of (Ta$_{1-x}$Mo$_x$)$_4$Si(Te$_{1-y}$Sb$_y$)$_4$. 
    The doping concentrations are given as $(x,y)=\{(0,0.05)$, $(0,0)$, $(0.001,0)$, $(0.002,0)$, $(0.01,0)$, $(0.02,0)$, $(0.05,0)\}$ from the front in (a) and from the left in (b), where $\Delta/W_c=\{0.055$, $0.04$, $0.05$, $0.04$, $0.005$, $0$, $0\}$ and $\lambda_0\times10^{3}=\{1,1,0.3,0.3,0.1,0.1,0.1\}$. 
    }
    \label{fig: doped_sigE}
\end{figure}

We also investigate the changes in the spectral conductivity and chemical potential of Ta$_4$SiTe$_4$ as the chemical doping is carried out, which are shown in Fig.~\ref{fig: doped_sigE}(a) and~(b).
For the hole-doped case or without any doping~$(x=0,y=0.05\textrm{ or }0.0)$, we find peaks for both conduction and valence bands that are separated by a band gap $\Delta_g\approx 0.15$~eV. 
Interestingly, the spectral conductivity is almost unchanged by hole-doping, while the chemical potential is shifted towards the valence band~[see Fig.~\ref{fig: doped_sigE}(b)]. In contrast, we find that the spectral conductivity is strongly modified upon electron-doping. Within the small doping regime (e.g., $x=0.001,0.002$), the spectral conductivity becomes greater with a larger band width. In electron-doped samples, the contribution from the valance band is much smaller than the conduction band, hence we find no peaks in the valence band. 
A further increase in $x$ leads to a decrease of $\sigma(E)$, which can be understood from the broadening of the spectral conductivity due to the impurity scattering. Electron-doping also results in the shift of the Fermi energy with a small change of $x$. This is expected from a small carrier number of undoped Ta$_4$SiTe$_4$. Once the Fermi energy is shifted towards the conduction band at $x=0.001$, the change in chemical potential becomes small with increasing $x$ except for the largest doping concentration at $x=0.05$.

\section{Discussion and Conclusion}
\label{sec: conc}
Let us comment on the limitations of our approach. First, the training of NN{s} requires thermoelectric transport data over a wide range of temperatures without a phase transition. 
Second, with the NN, it is difficult to reproduce the low-temperature divergence of electronic conductivity in extremely clean metals. This is because the linear interpolation used for Eqs.~\eqref{eq:L11a} and~\eqref{eq:L12a} results in large errors when $\sigma(E)$ exhibits a sharp peak near the Fermi energy. This limitation can be overcome if the low-energy structure of $\sigma(E)$ is known from theory. In this case, the NN can be used to estimate a small correction in $\sigma(E)$ from the low-energy model in addition to the chemical potential $\mu(T)$.

To relate the spectral conductivity to the response coefficients, we employed the Sommerfeld-Bethe relation in Eqs.~\eqref{L11} and~\eqref{L12}. However, this relation does not hold in some cases.
For example, an electron-phonon interaction gives rise to the phonon-drag mechanism, which cannot be explained by the Sommerfeld-Bethe relation~\cite{jonsonElectronphononContributionThermopower1990, matsuuraEffectPhononDrag2019}. For strongly correlated systems, the Sommerfeld-Bethe relation was shown to hold with a short-range interaction such as the Hubbard model~\cite{kontaniGeneralFormulaThermoelectric2003}, but not with a finite-range interaction~\cite{ogataRangeValiditySommerfeld2019}.
It is also possible to obtain anomalous terms beyond the Sommerfeld-Bethe relation by coupling electrons with other excitations such as excitons and (para)magnons~\cite{takaradaTheoryThermalConductivity2021a, zhengParamagnonDragHigh,matsuuraTheoryHugeThermoelectric2021,endoEffectParamagnonDrag2022}.
Nevertheless, the Sommerfeld-Bethe relation holds for a wide range of materials. In particular, it will be interesting to apply our method to strongly correlated systems with short-range interactions, whose electronic band structure is difficult to obtain using first principles calculations~\cite{imadaElectronicStructureCalculation2010}. 


\begin{figure}[t]
    \centering
    \includegraphics[width=\columnwidth]{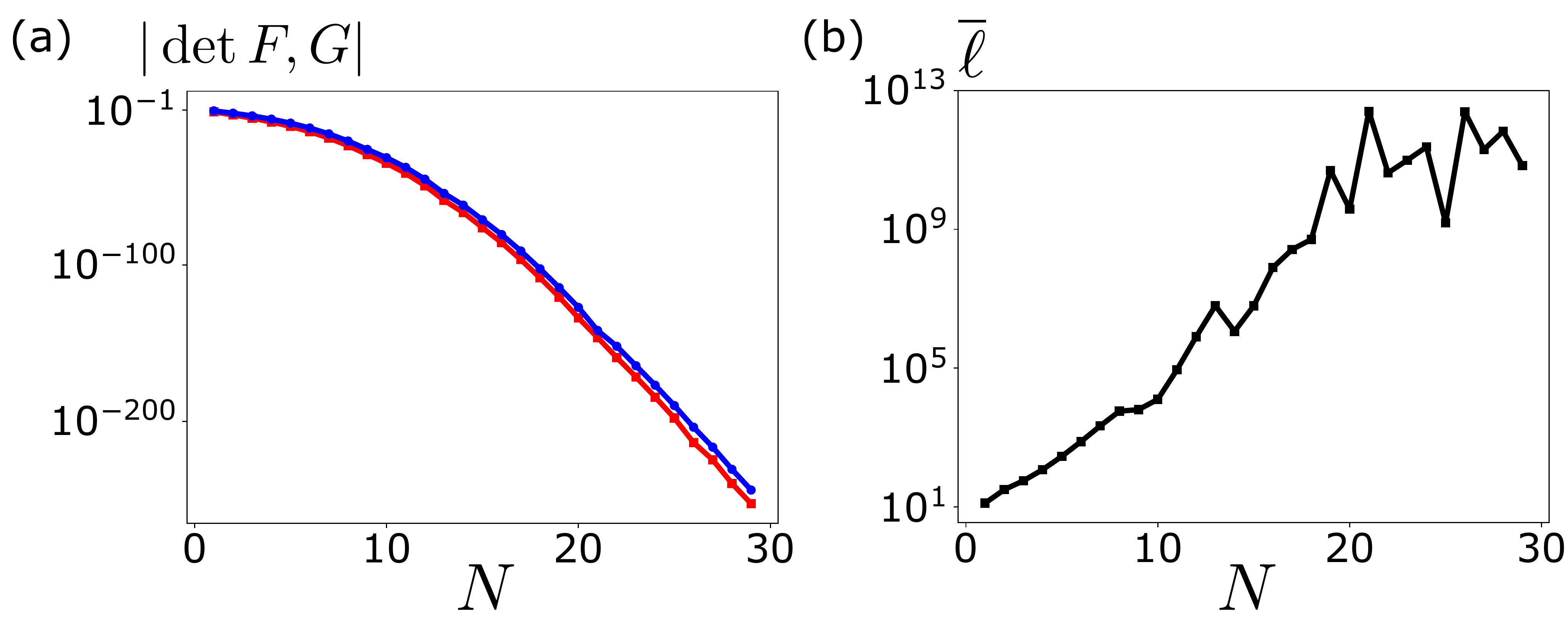}
    \caption{\textbf{Exponentially small determinants of $F$ and $G$ matrices result in large numerical errors.} (a)~The absolute value of determinants for $F$~(red with squares) and $G$ matrices~(blue with circles) are plotted against the matrix size~$N$. (b)~The normalized loss function~$\overline{\ell}$ is plotted against~$N$ using the toy model~$\sigma_\textrm{toy}(\CalE)$~[see Eq.~\eqref{eq: sigma_toy}]. The cutoff is taken as $W_c=5$ in both (a) and (b).
    }
    \label{fig: determinant_FG}
\end{figure}

In summary, we have proposed a data-driven approach to solve the inverse problem in thermoelectric phenomena using a NN. Based on the simple relationship between the thermoelectric coefficients and the spectral conductivity, our method enables the reconstruction of the spectral conductivity and chemical potential without prior knowledge about microscopic properties of target materials.
Applying our method to doped one-dimensional telluride Ta$_4$SiTe$_4$~\cite{inoharaLargeThermoelectricPower2017}, we have obtained the spectral conductivity and chemical potential at various doping concentrations. The reconstructed spectral conductivity was used to evaluate the electronic thermal conductivity $\kappa_\textrm{el}$ and the maximal figure of merit $ZT$. Crucially, our approach can provide accurate estimates of $\kappa_\textrm{el}$ and $ZT$ beyond the Wiedemann-Franz law, which is not valid in insulators and at high temperatures. Furthermore, the reconstruction of the complete energy dependence of the spectral conductivity, including the phenomenological relaxation time $\tau(E)$, opens a path to analyzing complex materials whose electronic states remain elusive. 
Therefore, this study provides deep insights into the connection between thermoelectric properties and the low-energy electronic states, and establishes a promising route to incorporate experimental data into traditional theory-driven workflows.

A python code of our method is provided at Ref.~\cite{hirosawaGit2022}.



\begin{acknowledgements}
We are grateful to A. M. M\"{u}ller and C. Bruder for discussions. T. H. is supported by Japan Society for the Promotion of Science (JSPS) through Program for Leading Graduate Schools (MERIT) and JSPS KAKENHI (Grant No. 18J21985). F. S. acknowledges financial support from the NCCR QSIT funded by the Swiss National Science Foundation (Grant No. 51NF40-185902). This work was supported by Grants-in-Aid for Scientific Research from the Japan Society for the Promotion of Science (No. JP18H01162, No. JP18K03482, No. JP20K03802, and No. JP21K03426), and the JST-Mirai Program Grant Number JPMJMI19A1, Japan.
\end{acknowledgements}

\appendix

\section{Exponentially small determinants of $F$ and $G$ matrices}
\label{section: determinant}

Let us define $F$ and $G$ as the following $N\times N$ matrices:
\begin{align}
    F(\CalE_j,\xi_i)&=
    \frac{\exp[\CalE_j/\xi_i]}{\xi_i (\exp[\CalE_j/\xi_i]+1)^2}, 
    \\
    G(\CalE_j,\xi_i)&
    =-\,\,\frac{\CalE_j \exp[\CalE_j/\xi_i]}{\xi_i (\exp[\CalE_j/\xi_i]+1)^2}, 
\end{align}
where dimensionless variables are defined as $\xi_i=T_i/T_N$ and $\CalE_j=W_c\xi_j$ for $i,j=1,2,\ldots, N$ with an integer $N$ and a cutoff $W_c$. From Eqs.~\eqref{eq: Mat_L11} and~\eqref{eq: Mat_L12}, the symmetric and antisymmetric parts of the spectral conductivity are given by $\sigma_\textrm{sym}(\CalE_j)=(2W_c/N)^{-1}\sum_{j=1}^NF(\CalE_j,\xi_i)^{-1}L_{11}(\xi_i)$ and $\sigma_\textrm{anti}(\CalE_j)=(2W_c/N)^{-1}\sum_{j=1}^NF(\CalE_j,\xi_i)^{-1}\overline{L}_{12}(\xi_i)$, respectively.
However, the determinants of $F$ and $G$ are exponentially small with the matrix size $N$, as shown in Fig.~\ref{fig: determinant_FG}(a). Here, the energy cutoff is fixed as $W_c=5$.
The exponentially small determinants of $F$ and $G$ result from the sharply peaked Fermi-Dirac distribution function.
Since inverse matrices are proportional to the inverse of their determinants, floating-point overflow occurs.
As a result, it is not possible to compute $F^{-1}$ and $G^{-1}$ using a numerical solver for inverse matrices.
Figure~\ref{fig: determinant_FG}(b) shows that the normalized loss function~$\overline{\ell}=\ell/N$ increases exponentially with the matrix size $N$, which is evaluated with $a_i=1$ and $b=0$ in Eq.~\eqref{eq: loss}. Test datasets for $L_{11}(\xi_i)$ and $\overline{L}_{12}(\xi_i)$ are generated from the toy model $\sigma_\textrm{toy}(\CalE)$~[see Eq.~\eqref{eq: sigma_toy}]. Although there is no floating-point overflow for a small number of $N$, the discretization error converting the integrals in Eqs.~\eqref{L11}-\eqref{L12} to the matrix products in Eqs.~\eqref{eq: Mat_L11}-\eqref{eq: Mat_L12} becomes significant as $N$ becomes smaller. This is the reason why $\overline{\ell}$ exceeds $10^1$ even for small values of $N$.

\section{Details of the data-driven
interpolating model}
\label{section: details_NN}
Our data-driven method consists of three steps. First, we need raw experimental data of the electric conductivity $\sigma(T)$ and Seebeck coefficient~$S(T)$ at multiple chemical doping concentrations. 
For the results presented in Sec.~\ref{section: result}, we used the method available in Ref.~\cite{Rohatgi2020} to extract numerical data points from the figures in Ref.~\cite{inoharaLargeThermoelectricPower2017}. The data is interpolated to prepare the linearly spaced temperature grids with corresponding values of $\sigma$ and $S$, which is then used to compute $L_{11}=\sigma$ and $L_{12}=TS\sigma$ as shown in Fig.~\ref{fig: exp_data}. Second~(presented in Sec.~\ref{sec: sigmaE_N=7}), the NN is employed to infer the functional form of the spectral conductivity $\sigma(E)$ and chemical potentials~$\mu^{(s)}(T)$ for $s=1,2,\ldots,N_s$. At this step, the spectral conductivity is assumed to be unchanged by chemical doping.
Third~(presented in Sec.~\ref{sec: sigmaE_optimized}), the spectral conductivity is optimized to reproduce $L_{11}(T)$ and $L_{12}(T)$ for a given doping concentration using $\mu^{(s)}(T)$. When the fitting is not satisfactory, we introduce a small correction $\delta \mu^{(s)}(T)$ in the chemical potential. In the following, we provide details of the NN architecture and the training procedure.

\subsection{Neural network architecture}
We use a fully connected NN with three layers.
The first layer acts as an input layer, taking the vector $\{\xi_i\}$ with $i=1,2,\ldots,N$. 
Except for the output layer, we apply rectified linear units as activation functions, defined as $\textrm{max}(0, z)$. No activation function is used in the output layer because it turns out to slow down or stop the learning process. Instead, we add a term to penalize the negativity of $\sigma(\CalE_j)$ in the loss function as defined in Eq.~\eqref{eq: loss} of Sec.~\ref{section: NN model}.

\subsection{Training procedure}
The NNs are implemented in PyTorch~\cite{NEURIPS2019_9015}.
For the training of the NN, the spectral conductivity is defined as $\sigma(\CalE_j)=\sum^{10}_{k=1}o_{j+k}/10$ for $j=-N,-N+1,\ldots, N$ $(N_s=1)$ or $j=-2N,-2N+1,\ldots, 2N$ $(N_s>1)$, with $o_j$ denoting the output of the third layer.
For $N_s>1$, $\mu(\xi_i)$ is also obtained from Eq.~\eqref{eq: chemical_model}. Using $\sigma(\CalE_j)$ and $\mu(\xi_i)$, we compute $L_{11}$, $L_{12}$, and the mean-squared error loss function, which is defined as the difference from the experimental values of $L_{11}$ and $L_{12}$ in Eq.~\eqref{eq: loss}. 
The loss function is minimized as the weights and biases are optimized by the stochastic gradient-based optimizer Adam~\cite{kingmaAdamMethodStochastic2017}. Gradients are calculated by backpropagation~\cite{rumelhartLearningRepresentationsBackpropagating1986}.
We train the NN with the full batch size, performing a conditional update of the learning rate:
When the loss function increases at the $i$th epoch~($\ell_i>\ell_{i-1}$), the learning rate at the $(i+1)$th epoch is updated as $\lambda_{i+1}=\textrm{max}(\lambda_i \cdot 0.8,\lambda_\textrm{min})$, where $\lambda_\textrm{min}=\textrm{min}(\lambda_0/5,\ell_{i})$.
This procedure prevents oscillations in $\ell$ and leads to smooth optimization of $\sigma(\CalE_j)$ and $\mu(\xi_i)$.

\bibliographystyle{apsrev4-1}


%
\end{document}